\documentclass[twocolumn,showpacs,preprintnumbers,amsmath,amssymb,superscriptaddress,prl]{revtex4}

\usepackage{amsmath}

\usepackage{amssymb}

\usepackage{graphicx}

\usepackage{dcolumn}

\usepackage{bm}

\usepackage{color}

\def\be{\begin{eqnarray}}

\def\ee{\end{eqnarray}}

\def\ba{\begin{array}}

\def\ea{\end{array}}

\begin{document}




\title{Asymmetric non-linear response of the quantized Hall effect}

\author{A. Siddiki} %
\affiliation{Center for NanoScience and Fakult\"at f\"ur Physik,
Ludwig--Maximilians--Universit\"at, Geschwister--Scholl--Platz 1,
D--80539 M\"unchen, Germany} \affiliation{Physics Department,
Faculty of Arts and Sciences, 48170-Kotekli, Mugla, Turkey}



\author{J. Horas} %
\affiliation{Center for NanoScience and Fakult\"at f\"ur Physik,
Ludwig--Maximilians--Universit\"at, Geschwister--Scholl--Platz 1,
D--80539 M\"unchen, Germany}

\author{D. Kupidura} %
\affiliation{Center for NanoScience and Fakult\"at f\"ur Physik,
Ludwig--Maximilians--Universit\"at, Geschwister--Scholl--Platz 1,
D--80539 M\"unchen, Germany}

\author{W. Wegscheider} %
\affiliation{Laboratory for Solid State Physics, ETH Z\"urich, CH-8093 Z\"urich, Switzerland}

\author{S. Ludwig} %
\affiliation{Center for NanoScience and Fakult\"at f\"ur Physik,
Ludwig--Maximilians--Universit\"at, Geschwister--Scholl--Platz 1,
D--80539 M\"unchen, Germany}

\begin{abstract}
An asymmetric break-down of the integer quantized Hall effect is
investigated. This rectification effect is observed as a function
of the current value and its direction in conjunction with an
asymmetric lateral confinement potential defining the Hall-bar.
Our electrostatic definition of the Hall-bar via Schottky-gates
allows a systematic control of the steepness of the confinement
potential at the edges of the Hall-bar. A softer edge (flatter
confinement potential) results in more stable Hall-plateaus,
i.\,e.\ a break-down at a larger current density. For one soft and
one hard edge the break-down current depends on the current
direction, resembling rectification. This non-linear magneto-transport effect confirms the
predictions of an emerging screening theory of the IQHE.
\end{abstract}




\maketitle

The discovery of the integer quantized Hall effect (IQHE)
\cite{vKlitzing80:494} in a two-dimensional electron system (2DES)
subject to a perpendicular magnetic field $B$ opened a wide
research field in solid state physics, which became a paradigm
since then \cite{Dassarma}. In spite of many experimental
\cite{Ahlswede01:562,Yacoby04:328,josePHYSE,Siddiki:EPL02} and theoretical
\cite{Laughlin81,Halperin82:2185,Buettiker86:1761} efforts our
understanding of the IQHE is still far from being complete. The
conventional theories
\cite{Laughlin81,Buettiker86:1761,Kramer03:172} can successfully
describe the main features of the IQHE, namely the existence of
extended Hall-plateaus and their extremely accurate quantized
resistance values. Relying on a single-particle picture they fail
to give a comprehensive description of many experimental
observations on a more detailed level. These original edge or bulk
theories disregard the classical Hartree-type (direct)
Coulomb-interaction within the 2DES altogether
\cite{Laughlin81,Halperin82:2185,Buettiker86:1761}. The bulk
theories assume the current to flow through the entire Hall-bar
\cite{Kramer03:172}. Based on material properties such as disorder
they predict localized states. The edge-theories describe the
Hall-plateaus by assuming current flow only along the edges of the
Hall-bar \cite{Buettiker86:1761}. An explanation of the transition
region between plateaus requires the additional assumption of
localized bulk states (as predicted within the bulk theories). A
newer approach disregards disorder but includes the direct
Coulomb-interaction between electrons moving in the confinement
potential in a \emph{non} self-consistent manner
\cite{Chklovskii92:4026}. Building on this model a screening
theory emerged. It is based on self-consistent calculations
\cite{Lier94:7757} and in addition considers disorder as well as
the quantum mechanical wave functions of the electrons
\cite{siddiki2004,Bilayersiddiki06:,SiddikiEPL:09}. For the
Fermi-energy approximately centered in between two adjacent
Landau-levels the screening theory predicts, that the current is
carried by incompressible regions (strips) extending along the
Hall-bar, hence replacing the edge-channels. Since back-scattering
is absent within an incompressible region this explains the
observation of the Hall-plateaus. Moreover, this self-consistent
approach allows predictions going beyond the scope of the
conventional theories, for example by taking into account the
exact shape of the confinement potential or explicitly considering
the non-linear transport regime. Effects based on the electron
spin such as exchange interaction are not taken into account here,
but can be included \cite{afifPHYSEspin}.

In this letter we discuss non-linear magneto-transport
measurements in the framework of the screening theory. We
experimentally investigate the current induced break-down of the
IQHE in gate-defined Hall-bars with laterally asymmetric
confinement potentials. In detail, we demonstrate a situation, in
which dissipation-less current only exists in one of the two
possible current directions. In agreement with the screening
theory this rectification of the IQHE occurs in a wide range of
parameters such as the mobility, charge carrier density and
Hall-bar width, as long as disorder effects do not dominate the
formation of incompressible strips \cite{SiddikiEPL:09}.

\begin{figure} {\centering
\includegraphics[width=1\linewidth]{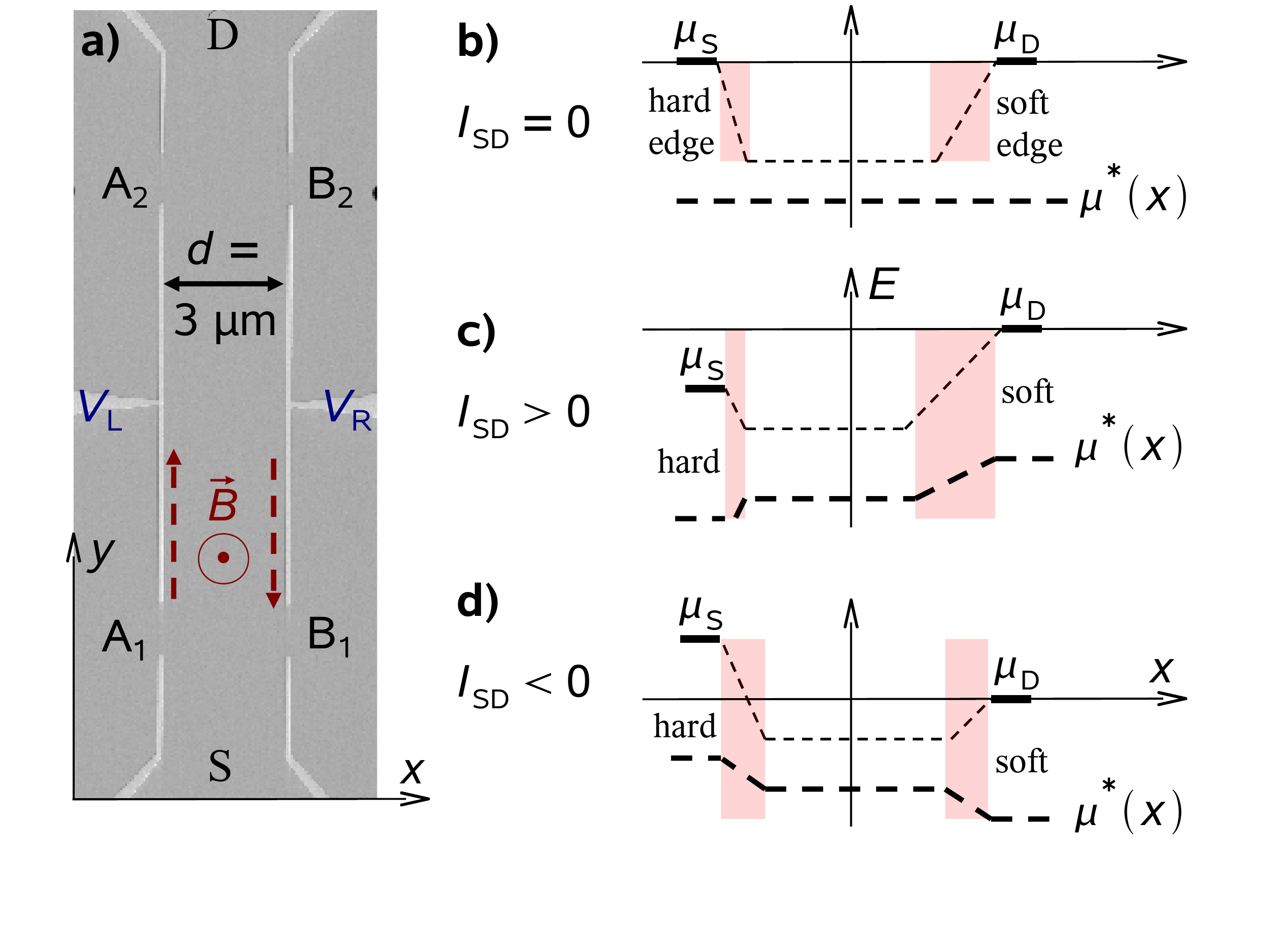}\vspace{-7mm}
\caption{ \label{fig1} (Color online) a) Scanning electron
microscope photograph of a typical sample. Top gates are colored
in light gray. The voltage $V_\mathrm L$  ($V_\mathrm R$) is
applied to the three lhs (rhs) gates. A constant current
$I_\mathrm{SD}$ is impressed at the source (S) contact and flows
into the grounded drain (D) contact. The other ohmic contacts
A$_1$, A$_2$, B$_1$ and B$_2$ are used as voltage probes. A
magnetic field perpendicular to the 2DES is directed upward and
defines a left-handed chirality for electrons moving along the
2DES (dashed arrows). b) Qualitative sketch of the energy of the relevant
Landau-level (thin dashed line), which is pinned to the chemical
potential (Fermi-energy) at the edges. Here $I_\mathrm{SD}=0$ and
$V_\mathrm L<V_\mathrm R<V_\mathrm{depl}<0$, where at
$V_\mathrm{depl}$ the 2DES beneath a gate is completely depleted;
$\mu_\mathrm S$ and $\mu_\mathrm D$ are the chemical potentials defined at
the source versus drain contacts. Also shown is the
electro-chemical potential $\mu^*(x)$ (fat dashed line) across the
Hall-bar. The shaded areas mark the width of incompressible
regions. c,d) Same as b) but for the non-equilibrium case $I_\mathrm{SD}>0$ or
$I_\mathrm{SD}<0$.}}
\end{figure}

Our Hall-bars are electrostatically defined by means of metallic
Schottky-gates produced by electron-beam lithography on the
surfaces of high mobility AlGaAs\,/\,GaAs-heterostructures
containing 2DESs 110\,nm beneath the surface. This field-effect
method allows to define Hall-bars with extremely smooth and
selectively tunable confinement potentials. A typical gate layout
is displayed in the SEM-picture of Fig.\ \ref{fig1}a.
A constant dc current is impressed between the source (S) and
drain (D) contacts, while four more ohmic contacts A$_1$, A$_2$
and B$_1$, B$_2$ are used as voltage probes. The Hall-resistance
$R_\mathrm H$ is obtained measuring the voltage drop between
contacts A$_1$ and B$_1$ or A$_2$ and B$_2$, while the
longitudinal resistance $R_\mathrm L$ is measured with A$_1$ and
A$_2$ or B$_1$ and B$_2$. For simplicity we will not specify which of these combinations of contacts are used in the following, given that our measurements are
roughly independent of it. In order to create a laterally
asymmetric confinement potential we apply different gate voltages
$V_\mathrm L$ and $V_\mathrm R$ along the two sides of the
Hall-bar, while the three gates on each side are always on equal
potential. In all measurements shown here $B$ is perpendicular to
the 2DES and points upwards, thus defining left-handed chirality,
as sketched in Fig.\ \ref{fig1}a for the linear response case
(dashed arrows indicate the direction that electrons move in
equilibrium). $I_\mathrm{SD}>0$ corresponds to $V_\mathrm S>0$
while the drain contact is always grounded $V_\mathrm D=0$ (for the measurements shown in this letter). Figs.\
\ref{fig1}b, \ref{fig1}c and \ref{fig1}d sketch the energy of the
relevant Landau-level (thin dashed line) for $I_\mathrm{SD}=0$
(Fig.\ \ref{fig1}a), $I_\mathrm{SD}>0$ (Fig.\ \ref{fig1}b) and
$I_\mathrm{SD}<0$ (Fig.\ \ref{fig1}c) as predicted by the
screening theory for the Hall-plateaus \cite{SiddikiEPL:09}. The
relevant Landau-level is the one, which is pinned to the chemical
potentials at the edges of the Hall-bar. Also shown is the
electro-chemical potential $\mu^*(x)$ (fat dashed line) across the
Hall-bar, which includes the effect of a non-zero current, i.\,e.\
the non-equilibrium case. The shaded areas mark the width of
incompressible regions and will be discussed below.

We have performed measurements on five different samples with
Hall-bar width of 3\,$\mu$m or 10\,$\mu$m and mobilities of
$\mu\simeq1.4\,\times\,10^6\,\mathrm{cm}^2\,/\,\mathrm{Vs}$,
$\mu\simeq3\,\times\,10^6\,\mathrm{cm}^2\,/\,\mathrm{Vs}$, and
$\mu\simeq8\,\times\,10^6\,\mathrm{cm}^2\,/\,\mathrm{Vs}$ on the
wafers I, II, and III, respectively. Here, we only present data
measured on wafers I and III at a temperature of $T\simeq 1.7\,$K.
However, all our data taken so far confirm the results discussed
below.

\begin{figure} {\centering
\includegraphics[width=1\linewidth]{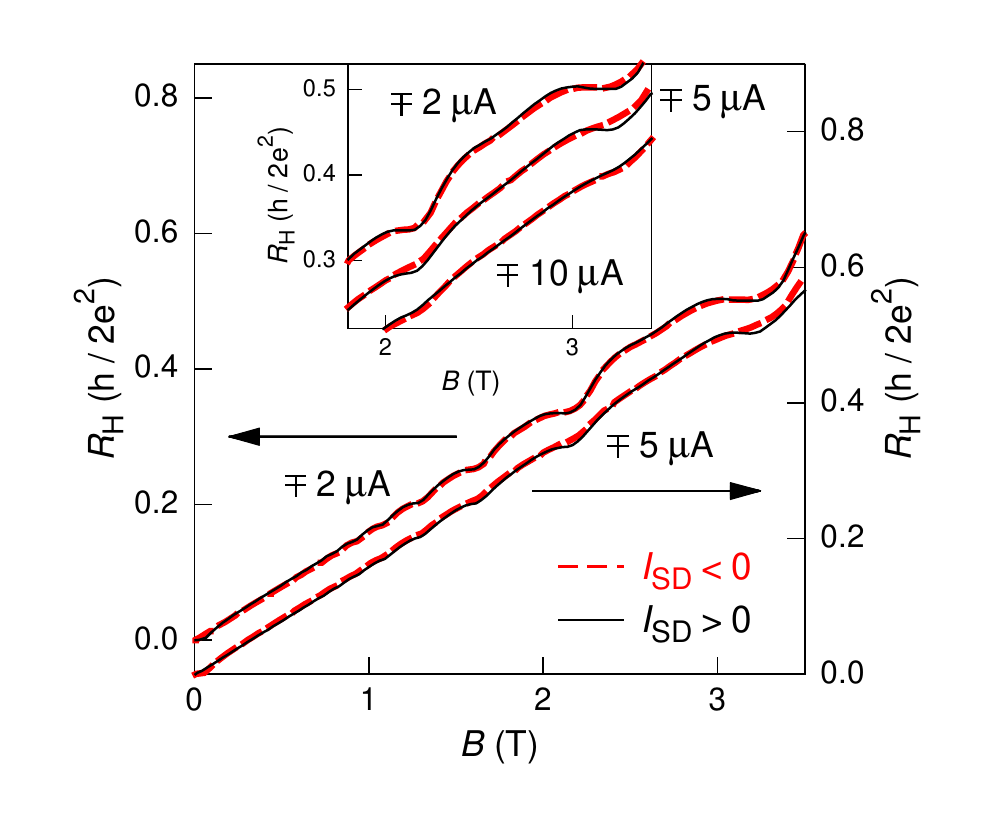}
\caption{ \label{fig2} (Color online) Hall-resistance of a
Hall-bar with a width of $d=10\,\mu$m defined in wafer I
($\mu\simeq1.4\times10^6\,\mathrm{cm}^2\,/\,$Vs). An asymmetric
confinement potential is created with $V_\mathrm L=-1.2\,$V and
$V_\mathrm R=-0.3\,$V causing a hard lhs edge and a soft rhs edge
of the Hall-bar. The main plot shows $R_\mathrm H$ for
$I_\mathrm{SD}=\mp2\,\mu$V (lhs y-axis) and
$I_\mathrm{SD}=\mp5\,\mu$V (rhs y-axis). The inset displays
detailed views of the section including filling factors $\nu=6$
and $\nu=4$ of the same data and, in addition, for
$I_\mathrm{SD}=\mp 10\,\mu$A. For clarity the curves for
$|I_\mathrm{SD}|> 2\,\mu$A in the inset are vertically shifted.}}
\end{figure}

Fig.\ \ref{fig2}
displays the measured Hall-resistance $R_\mathrm H$ of a Hall-bar realized in wafer I
($d=10\,\mu$m, $\mu\simeq1.4\,\times\,10^6\,\mathrm{cm}^2\,/\,\mathrm{Vs}$) as a function
of the perpendicular magnetic field $0<B<3.5\,$T. The gate voltages are $V_\mathrm
L=-1.2\,$V and $V_\mathrm R=-0.3\,$V. They create a harder confinement potential on the
left hand side (lhs) of the Hall-bar compared to the relative soft rhs edge (see Fig.\
\ref{fig1}b). Note, that $V_\mathrm{depl}=-0.3\,$V just allows complete depletion of the
2DES beneath any of the biased gates. The Hall-curves displayed in Fig.\ \ref{fig2} are
measured at $I_\mathrm{SD}=\mp 2\,\mu$A  (lhs y-axis) and $I_\mathrm{SD}=\mp 5\,\mu$A (rhs
y-axis). Part of the curves, namely in the region of filling factors $\nu\simeq6$ and
$\nu\simeq4$ are shown again in the inset of Fig.\,\ref{fig2}, where we also added data
for $I_\mathrm{SD}=\mp 10\,\mu$A (curves for $I_\mathrm{SD}=\mp 5\,\mu$A and
$I_\mathrm{SD}=\mp 10\,\mu$A are vertically shifted). For $I_\mathrm{SD}=\mp 2\,\mu$A the
Hall-plateaus are well established, while they are pretty much smeared out for
$I_\mathrm{SD}=\mp 10\,\mu$A, independent of the current direction. This observation can
be attributed to the well known break-down of the Hall-effect, usually explained (within
the edge channnel models) by scattering between (many) edge channels
\cite{Guven02:155316}. Interestingly, for the intermediate current value of
$I_\mathrm{SD}=\mp 5\,\mu$A the break-down is more pronounced for one of the two current
directions, namely $I_\mathrm{SD}<0$.

Exactly this behavior is predicted by the screening theory \cite{SiddikiEPL:09}; within
this calculation scheme the current induced break-down of the Hall-effect is caused by
inelastic scattering between compressible regions (Joule heating) \cite{Akera06:}.
Essentially the width of the incompressible strips decreases as the current is increased.
However, as long as there exists at least one incompressible strip across the Hall-bar,
dissipation less current is possible resulting in the plateau-value of $R_\mathrm H$. In
agreement with the screening theory we assume two incompressible strips, just one on each
edge of the Hall-bar \cite{siddiki2004}. On the one hand, the asymmetric confinement
causes the incompressible strip on the softer edge to be wider than the one on the harder
edge (for Fig.\ \ref{fig2} the lhs edge), as sketched in Fig.\ \ref{fig1}b. On the other
hand, a large current generally results in a widening (narrowing) of the incompressible
strip at the edge of the higher (lower) electrochemical potential (Figs.\ \ref{fig1}c and
\ref{fig1}d) \cite{SiddikiEPL:09}. For the higher electrochemical potential on the softer
edge (Fig.\ \ref{fig1}c), the result is therefore a very narrow incompressible strip on
the hard edge and a very wide incompressible strip on the soft edge. The direct
consequence is a more stable incompressible strip (on the soft edge) and a wider
Hall-plateau at the onset of the break-down regime. For the data shown in Fig.\ \ref{fig2}
this situation is reached for $I_\mathrm{SD}>0$. For the opposite current direction $I_\mathrm{SD}<0$ the
two effects (namely current versus asymmetry induced widening or narrowing of
incompressible strips) cancel out and, accordingly, the break-down is already observed
for lower absolute values of the current (Fig.\ \ref{fig1}d).

\begin{figure} {\centering
\includegraphics[width=1\linewidth]{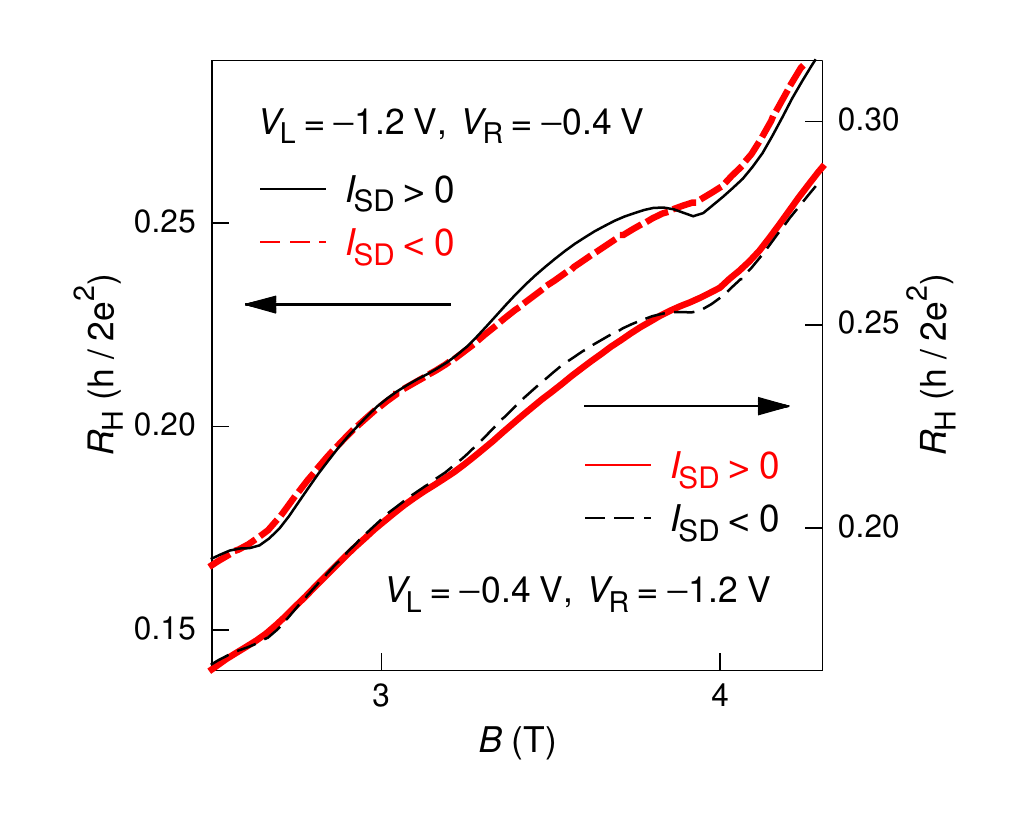}
\caption{ \label{fig3} (Color online) Hall-resistance of the same
Hall-bar as for Fig.\,\ref{fig2}
($d=10\,\mu$m, $\mu\simeq1.4\times10^6\,\mathrm{cm}^2\,/\,$Vs) for
$I_\mathrm{SD}=\mp6\,\mu$m. In addition to the current direction
the lateral asymmetry is modified exchanging the soft and hard
edges. These data where measured after illumination of the sample
at $T=1.7\,$K, causing a higher charge carrier density but no
qualitative change on the investigated effects.}}
\end{figure}

Within this scenario it is possible to compensate a reversal of the lateral asymmetry of
the confinement potential by reversing the direction of the impressed current. This
prediction \cite{SiddikiEPL:09} is experimentally tested in Fig.\,\ref{fig3},
where we plot $R_\mathrm H\,(B)$ taken at $I_\mathrm{SD}=\mp6\,\mu$m on the same wafer as
above. Between the two sets of curves the asymmetry of the confinement potential is
reversed by exchanging the soft and hard edges of the Hall-bar (while the magnetic field
direction is unchanged). For the harder edge on the lhs of the Hall-bar (lhs axis, two
curves on top) the Hall-effect is more stable for $I_\mathrm{SD}>0$, as already observed
in Fig.\,\ref{fig2}. In contrast, if the lhs edge is the softer one, the Hall effect is
more stable for $I_\mathrm{SD}<0$. As expected the general behavior stays unchanged,
whenever we inverse both, the direction of the current and the direction of the lateral
confinement potential. However, a reversal of only one of the two quantities near the
onset of the break-down of the IQHE causes a drastic change in the width of the
Hall-plateaus. We interpret this behavior as a rectification of the IQHE.

\begin{figure} {\centering
\includegraphics[width=.95\linewidth]{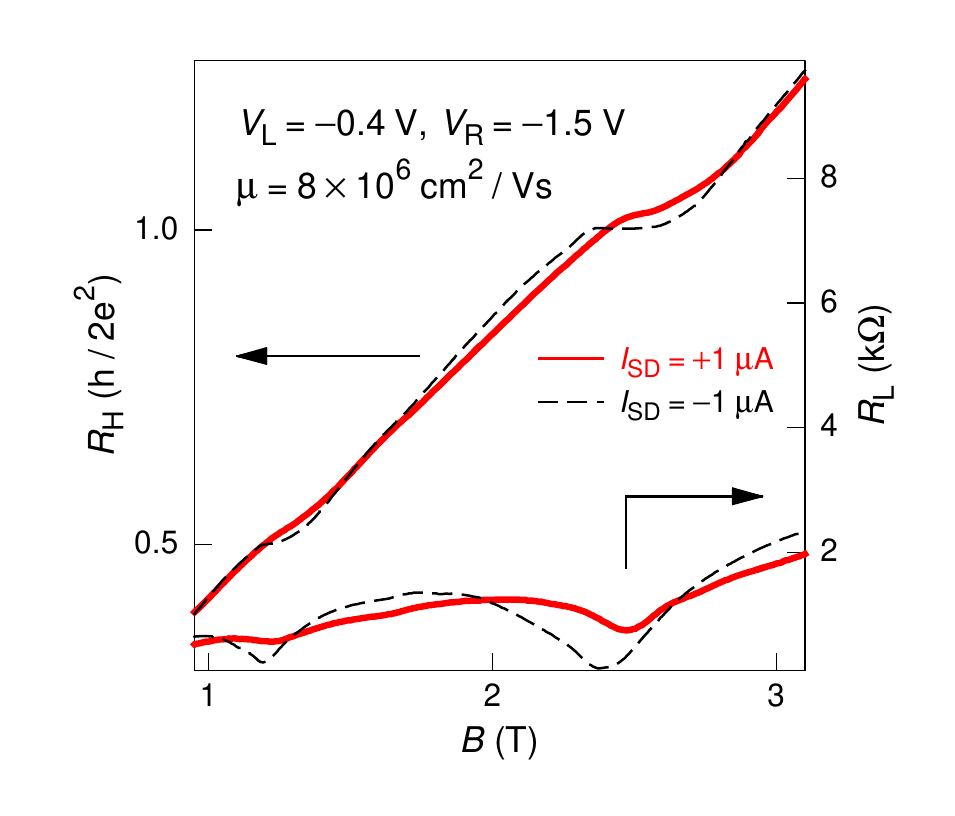}
\caption{ \label{fig4} (Color online) Hall-resistance $R_\mathrm
H$ of an asymmetric Hall-bar with a width of $d=3\,\mu$m on wafer
III ($\mu\simeq8\times10^6\,\mathrm{cm}^2\,/\,$Vs) for
$I_\mathrm{SD}=\mp1\,\mu$m. In addition shown are the
corresponding longitudinal resistances $R_\mathrm L$ (rhs
y-axis).}}
\end{figure}

Fig.\ \ref{fig4}
plots an example of the same behavior, but observed on wafer III with a much higher
mobility of $\mu\simeq8\,\times\,10^6\,\mathrm{cm}^2\,/\,\mathrm{Vs}$ and a Hall-bar width
of $d=3\,\mu$m measured at $I_\mathrm{SD}=\mp1\,\mu$A. Here we apply $V_L=-0.4\,$V and
$V_R=-1.5\,$V defining the harder edge on the rhs of the Hall-bar. Note, that the smaller
absolute value of the break-down current observed in the high mobility sample might be
explained by a smaller width of this Hall-bar. As expected the break-down of the
Hall-effect is more pronounced for positive current, causing a higher chemical potential
at the lhs edge of the Hall-bar. In Fig.\ \ref{fig4} we additionally display the
longitudinal resistance (rhs y-axis). It shows a break-down behavior in accordance to the
$R_\mathrm H\,(B)$ data.

We observe the same behavior in a wide range of mobilities, charge carrier densities,
Hall-bar widths and contact combinations. While here we show only a small selection of our
data, we have performed many control measurements always leading to the same systematic
result, namely rectification of the IQHE in a Hall-bar with an asymmetric lateral
confinement potential as a function of the direction of the impressed current (at the
onset of the break-down of the IQHE).

In the following we discuss our results in light of conventional theories versus the
screening theory of the IQHE. Both, the original bulk and edge theories fail to describe
the experimentally observed smooth transition regions between the plateaus of the
Hall-resistance (and the corresponding finite longitudinal resistance) in a self contained
way. Instead the transition between Hall-plateaus is phenomenologically explained by
assuming broadening of the Landau-levels and corresponding narrowing of the Hall-plateaus.
Hence, one would expect a narrowing of the Hall-plateaus as the mobility is decreased (and
the disorder is increased). However, in experiments the opposite behavior is observed,
namely the plateaus become wider as the mobility is decreased. To heal this discrepancy it
is then --again phenomenologically-- assumed, that disorder induced localization causes an
insulating bulk state. The edge theories take, in addition, the confinement potential into
account, as the edge states are a direct result of the Landau-levels cutting the
Fermi-energy at the edges of the Hall-bar. However, no detailed assumptions are made
regarding the shape of the confinement potential.

In our experiments, we observe the effect of an impressed current on the Hall-resistance
(altering the transition regions between the plateaus) as a function of the lateral shape
of the confinement potential. In this regime, we cannot expect the conventional theories
to explain our findings. Moreover, calculations within these conventional models are done
in the linear response regime, while here we use large currents clearly putting us out of
the linear response regime.

The screening theory is based on numerical calculations of the exact shape of the
confinement potential by taking the direct Coulomb-interaction between charge carriers as
well as their quantum mechanical properties into account \cite{siddiki2004,SiddikiEPL:09}.
In contrast to the conventional theories, the screening theory allows self-consistent
numerical calculations even in the non-linear response regime, which results in
predictions at the onset of the break-down regime of the IQHE. The observed rectification
of the IQHE at the onset of its break-down go beyond the scope of conventional theories.
Our results qualitatively confirm the predictions of the screening theory.

It should be noted that the present form of the screening theory does not
account for the local temperature (and local heating effects) in a self-consistent way.
However, a reasonable prescription for such a calculation is already given in the
literature \cite{Akera06:}. In this calculation scheme a large impressed current
melts (narrows) the incompressible strips, finally leading to the experimentally observed breakdown. While the screening theory predicts the width and location of incompressible strips omitting local heating effects, the additional assumption of local heating as treated in reference \cite{Akera06:} results in a qualitative agreement with our experiments.

A.\,S.\ thanks D.\ Harbusch and D.\ Taubert for their technical support and J.\,P.\
Kotthaus for trusting a theorist performing experiments in his institute. Financial
support by the German Science Foundation via SFB 631, the Germany Israel program DIP and
the German Excellence Initiative via the "Nanosystems Initiative Munich (NIM)" is
gratefully acknowledged.




\end{document}